# Spatial and temporal distribution of 1 MeV proton microbeam guided through a poly(tetrafluoroethylene) macrocapillary

G. U. L. Nagy[1,2,*], Z. T. Gaál[2], I. Rajta[1], K. Tőkési[1]

[1]*Institute for Nuclear Research (Atomki), Bem square 18/c, H-4026 Debrecen, Hungary*
[2]*Institute of Physics, University of Debrecen, Bem square 18/a, H-4026 Debrecen, Hungary*

We present computer simulations about the spatial and temporal evolution of 1 MeV proton microbeam transmitted through an insulating macrocapillary with the length of 45 mm and with the inner diameter of 800 μm. The axis of the capillary was tilted to 1° relative to the axis of the incident beam, which ensured geometrical non-transparency. The simulation is based on the combination of stochastic (Monte-Carlo) and deterministic methods. It involves 1) random sampling of the initial conditions, according to distributions generated by the widely used and freely available computer software packages, SRIM and WinTRAX, 2) the numerical solution of the governing equations for following the classical trajectory of the projectiles, and 3) the description of the field-driven charge migration on the surface and in the bulk of the insulator material. We found that our simulation describes reasonably all of our previous experimental observations, indicating the functionality and reliability of the applied model. In addition, we found that at different phases of the beam transmission, different atomic processes result in the evolution of the beam distribution. First, in a scattering phase, the multiple small angle atomic scattering dominates in the beam transmission, resulting an outgoing beam into a wide angular range and in a wide energy window. Later, in a mixed phase, scattering and guiding happens simultaneously, with a continuously increasing contribution of the guiding. Finally, in the phase of the stabilized, guided transmission, a quadrupole-like focusing effect is observed, i.e. the transmitted beam is concentrated into a small spot, and the transmitted protons keep their initial kinetic energy.

## I. INTRODUCTION

When an incident ion beam enters an insulating capillary, whose axis is sufficiently tilted with respect to the axis of the ion beam, it will hit the inner surface of the capillary, and will deposit their electric charge due to ion-surface in teraction. Consequently, a charge patch forms on the inner capillary surface, generating an electric field which will in teract with subsequently arriving ions, and will deflect them towards the exit of the capillary. If the accumulated charge on the inner capillary wall, and thereby the electric field, is sufficiently high, the subsequently arriving ions can avoid the close collision with the surface. The accumulated charges can also interact with each other, and the charge-up of the inner 39 capillary wall in a self-organized manner occurs. In the lack 40 of close collisions between projectile ions and the atoms of the capillary material, the incident ion beam can exit the capillary with its initial kinetic energy and charge state. This phenomenon is called ion guiding. It was observed by Stolterfoht *et al.* [1], using a bunch of nanocapillaries etched in a polymer foil, and 3-keV Ne[7+] ions. Since then, numerous experimental and theoretical works, often motivated by potential technical applications, have been carried out to understand the complex nature of ion guiding. In recent years, these works have been summarized in several review papers [2–4].

Practical exploitation of the ion guiding phenomenon is almost restricted to the focusing of ion beams primarily in the MeV energy range, using tapered capillary optics. The transmission of MeV ions through insulating capillaries is mainly attributed to multiple small angle forward scatterings. This might have implied undesired changes, for example, energy loss and energy broadening of the beam. Nevertheless, ion beam focusing by means of tapered glass capillaries has become quite a well-established technique [5]. Despite that pure electrostatic guiding or focusing of MeV ions with capillaries of various shapes has still many open questions to solve, several laboratories built their home-made "capillary setup" for focusing ion beams for ion-beam analysis [6], for ion beam modification of materials [7], or for ion irradiation of living cells [8]. Development of these setups for new applications is continuously under way [9,10], nevertheless, to overcome experimental difficulties, it is also important to continue fundamental research, which reveals underlying processes and thus helps us in the design and construction of operational devices.

Computer simulations are powerful tools of research, and they have always played an important role in the field of ion guiding studies. Although the simulations are always based on simplified models, they can, at least in a given parameter range, describe experiments and predict new possible observables. During the past years several milestones in the theoretical description of ion guiding were achieved through computer simulations. In most cases good agreement was obtained between experiments and theory. Just a few examples are the following: (i) the oscillatory behavior of ion guiding, which was experimentally observed by Stolterfoht et al. [11], had already been predicted by Monte Carlo simulations [12]; (ii) the role of the capillary shape on the focusing capability against the blocking property was already suspected in Ref. [13], and then experimentally confirmed in Ref. [14].

*Corresponding author: nagy.gyula@atomki.mta.hu

In order to gain understanding of the governing processes of MeV ion beam transmission, we started the investigation of the guiding possibility of a 1-MeV proton beam through a single, cylindrical insulating macrocapillary in 2012. After the construction of our experimental setup [15], we proved in a series of work [16–19], that by using suitable conditions, pure electrostatic guiding of the swift ion beam is indeed possible and observable. These conditions cover a range of parameters, including (i) a material, which has exceptionally high electrical resistance, to be able to store a large amount of electric charge for a long time, (ii) a proper geometry of the capillary target, i.e., large wall thickness to reduce leakage currents through the bulk material, and length to reduce leakage currents along the surface, and (iii) a proper geometrical arrangement, i.e., large distance between the primary charge patch and the electrically grounded electrodes, to increase resistivity. Here we note that we used a focused, microsized beam, since it was expected that it is much easier to obtain an efficient transmission and thus it is easier to detect and characterize. Later we started the construction of a two-dimensional computer simulation [20,21], which, despite its simplicity, well reproduced our experimental results and, even more, it was capable to give predictions for observables. Some of them were experimentally confirmed later [18].

Recently, Liu et al. investigated the transmission features of a 1-MeV $H^+$ beam through a straight, cylindrical insulating macrocapillary, using their own computer simulation code. Beyond the good reproduction of our main conclusions using a microfocused beam, they pointed out the importance of beam divergence [22]. They also investigated the transmission of a parallel macrosized beam, and found partial transmission, but still partly due to electrostatic deflection. In a next step they comprehensively studied the tilt angle dependence, and found that by increasing the tilting angle, scattered transmission starts to dominate over the ion guiding [23]. The very good agreement between our work and an entirely different approach gave us confidence to continue our research along this line.

The motivation of the present work was to further develop our existing simulation code with some more accurate and versatile solutions, and to extend it to three dimensions, in order to obtain a computer simulation which is able to reproduce experiments and thus which allows us to give a proper theoretical description of the guiding of the MeV proton beam. The three-dimensional simulation is able to take into account the curvature of the capillary wall, giving us deeper insight both into the charge patch formation on the capillary wall and the into the trajectory of the projectile particles. After we present in detail the simulation code we show results of the transmission features of the 1-MeV microfocused proton 136 beam. We report the spatial distribution of the transmitted beam, its relation with the state of the transmission process, and its temporal evolution.

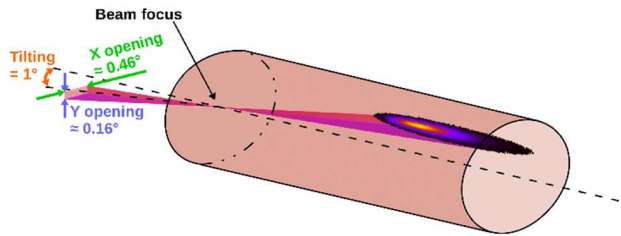

FIG. 1. Schematic illustration of the formation of the primary charge patch, according to the beam and target parameters used in this work. Note that the capillary is compressed along the longitudinal axis for better visibility.

## II. STRUCTURE OF THE SIMULATION

The simulation contains (1) stochastic processes, for which random numbers are generated using Monte Carlo methods. These processes involve the generation of the incident beam coordinates and angles, and the generation of the energy loss and scattering angle data upon collisions with the capillary wall, and (2) deterministic processes, in which the governing differential equations are solved numerically. These processes involve the motion of the projectile protons and the charge flow in the capillary wall, in both of which the Coulomb potential is the known interaction potential.

### A. Parameters

The parameters used in the present work are the same as those used in all of our previous works, including experiments and simulations. The incident beam is a 1-MeV $H^+$ beam, focused down to a few μm (the actual beam size depends on the beam intensity; see later). The energy spread of the beam is 1.5 keV [full width at half maximum (FWHM)], according to calibration measurements using nuclear resonance reactions. The focus point of the beam is on the entrance plane of the capillary target, at the center of the entrance hole. The full opening angle of the focused beam is ≈0.46° in the horizontal (target rotation) plane and ≈0.16° in the vertical plane. The capillary target is made of poly(tetrafluoroethylene) (PTFE), or as it is often called, Teflon. The length of the capillary is $l$ = 45 mm, its inner diameter is $d$ = 800 μm, giving an opening angle of ≈1°, and its wall thickness is $t$ = 400 μm. The tilting angle (the angle between the beam axis and the capillary axis) of the target is also 1°, but thanks to the center injection of the beam, these parameters fully satisfy the geometric nontransparency, as the primary charge patch will accumulate at around the middle of the capillary (see schematically in Fig. 1). This charge patch will be primarily responsible for the deflection and focusing of the incident ion beam.

### B. Incident beam generation: WINTRAX

WINTRAX [24], an ion-optics ray-tracing software dedicated primarily to ion micro- and nanoprobes, was used to generate distributions for the initial position and direction of incoming particles. First, the nuclear microprobe system (object and collimator slits, focusing quadrupole lenses, drift spaces) of Atomki was set up in WINTRAX, and the excitation of the focusing quadrupole lenses was optimized to achieve the smallest possible beam radius (point focus). Then, a large number (one million) of 1-MeV protons were traced, and two-dimensional bitmap images of the X-Y, X-Θ, and Y-Φ phase spaces at the focus point were generated for typical beam intensities (i.e., typical object slit openings), as seen in Fig. 2. The brightness values of these gray-scale images served as discrete values of the probability density function of the desired lateral and angular distributions. Then, X and Y coordinates, and the corresponding Θ and Φ angles were generated from these images using a hit-or-miss random value generation method. WINTRAX can be, however, used to generate other types of beams (defocused, only collimated, etc.), without making any assumptions about their distributions (i.e., perfectly parallel, Gaussian, etc.), giving us versatility in a realistic random sampling of the initial conditions.

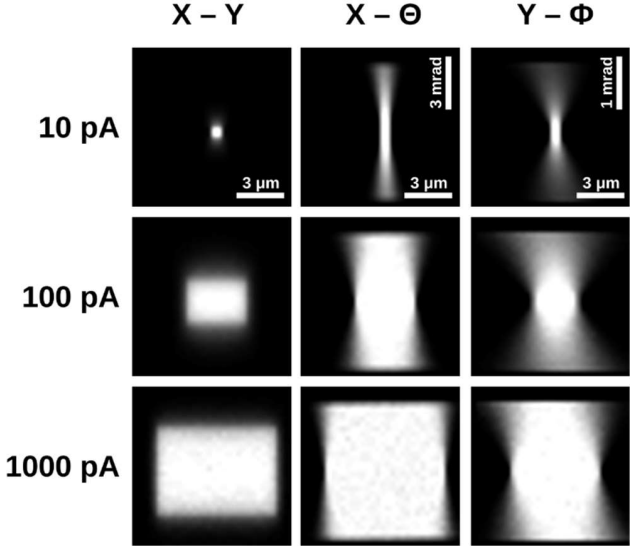

FIG. 2. Lateral and angular distribution of the incoming beam for typical beam intensities (slit openings), generated by WINTRAX. Scales in all columns apply to all intensities. It is seen that by increasing the beam intensity only the geometric beam size increases [(1) column vertical and horizontal, (2) and (3) column horizontal axis], the beam divergence does not increase [(2) and (3) column vertical axis].

### C. Trajectory of projectiles: Newton's equation of motion

The incident beam is treated as a series of individual particles, since space charge effect in this beam intensity range can be considered negligibly small. The trajectory of each projectile ion is approximated by solving Newton's equation of motion,

$$m\frac{d^2\vec{r}}{dt^2} = q\vec{E}, \qquad (1)$$

third-order velocity-Verlet algorithm. Here, $\vec{E} = (E_x, E_y, E_z)$ is calculated as the sum of the electric fields of point-like charges, $E(\vec{r}, t) = k \cdot \sum_{i=1}^{n} \frac{Q_i}{(\vec{r}_i - \vec{r})^2}$. The electric field of charges placed either on the surface or in the bulk of an insulator material is lowered due to dielectric screening [25], which depends on the relative dielectric constant of the material.

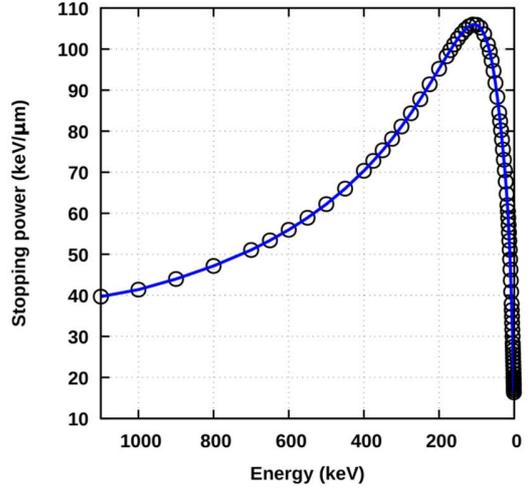

FIG. 3. Stopping power of proton beam in PTFE, calculated by SRIM.

### D. Energy loss and scattering of protons in material: SRIM

High energy light ions can deeply penetrate into bulk materials. For 1-MeV protons, according to the SRIM simulations, the penetration depth in PTFE is approximately 16.5 μm. The path of ions is rather straight, and the deposited energy per unit length is slowly increasing, except for the end of range, in the so-called Bragg peak, where the energy transfer has a maximum, and where the larger angle scatterings (scatterings on target nuclei) become important. In order to have a realistic, and simple approximation for this behavior, we used the stopping and range of ions in matter (SRIM) software package [26] to generate energy-loss (stopping power) and scattering angle data for our capillary simulation. The energy loss values are calculated at well defined energies by SRIM, as it can be seen in Fig. 3.

Using the energy values in Fig. 3, the transmission of a large number of protons through a thin PTFE foil was modeled by SRIM. The direction of each transmitted particle can be written into text files. The distribution is assumed to be Gaussian, whose mean value is 0º (i.e., transmission without change in direction) due to the radial symmetry, and the standard deviation is determined by Gaussian function fit. An example is shown in Fig. 4. It can be seen that at high energy the fit is not perfect, because at high energy the beam takes part in a low number of interactions while it traverses a thin foil. However, the difference is negligible, therefore in the following we will use the fitted distribution.

By collecting the standard deviations of all incident beam energies, we have a series of energy dependent scattering angle data (see Fig. 5). In our capillary simulation, these data points were used as input parameters. When an incident ion hit the capillary inner surface, the propagation into the bulk material is started. In a number of small steps the energy loss and scattering angle of the particle were calculated for every step by the interpolation between neighboring data points of Figs. 3 and 5. If the proton changed its direction in such a way that it could leave the material, the proton continues to travel with its actual, reduced kinetic energy, according to Eq. (1). However, if a proton lost its energy fully during its motion in the bulk material, a positive charge is deposited at this position.

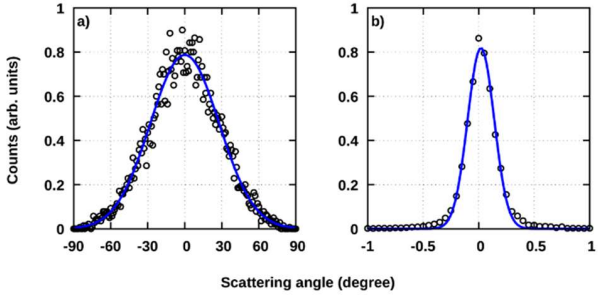

FIG. 4. Scattering angle of (a) 5-keV and (b) 1000-keV proton beam traversed through a 0.1-μm-thick PTFE foil. Open circles: SRIM simulation; solid line: Gaussian fit of the SRIM data.

SRIM simulations also revealed that in case of 1° (grazing) incidence, the mean value of the depth of deposited protons is only a few hundred nanometers below the surface (see Fig. 6). Since the equilibrium charge state for 1-MeV protons is close to +1 along most of their track, the charge deposition occurs most likely at the end of range. Here we note that, although the charge is deposited a few hundred nanometers below the surface, the model does not distinguish between bulk and surface. The capillary wall is taken into account as semi-infinite material.

**E. Field-driven charge redistribution: Nernst-Planck equation**

The key process in the formation of the successful and stable guided transmission is always the presence of the proper combination of charging and discharging mechanisms. While charging happens only by the incoming ion beam, there are several ways of discharging. Charge diffusion might dominate in the case of low electric fields, but, in the case of larger fields, the field-driven charge drift becomes more important. Furthermore, both can happen along the inner surface or through the bulk material to the outer surface, with different weight, because (i) the charge mobilities or conductivities on the surface and in the bulk differ from each other, and (ii) the size of the patch (the ratio of the side length to the area, through which surface and bulk currents, respectively, flow) also affects it. In addition to this, the charge neutralization by secondary electrons might also contribute to the discharge of the patch.

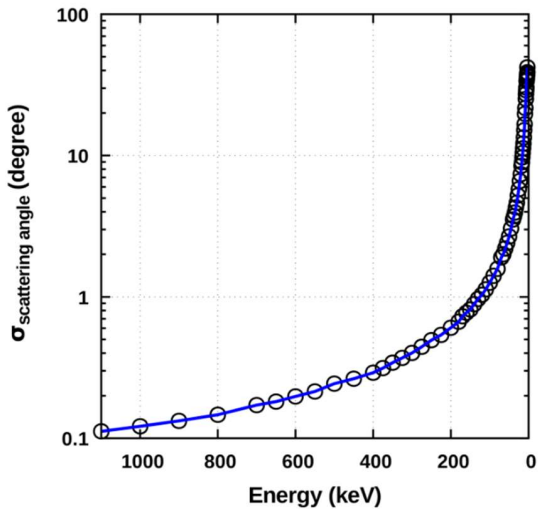

FIG. 5. Standard deviation of the scattering angle of proton beam traversed through a 0.1-μm-thick PTFE foil. Results of SRIM simulations.

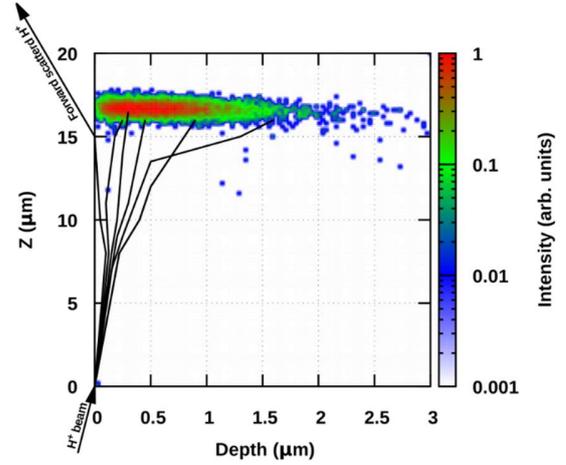

FIG. 6. Depth distribution of deposited protons in PTFE, in case of 1° (grazing) incidence, calculated by SRIM. The beam enters the material at the (0,0) position. Ion trajectories are shown to guide the eye.

In the present model, both the concentration gradient induced charge diffusion and the potential gradient induced charge drift were considered. The discharge mechanism was treated using the Nernst-Planck equation [27],

$$J(r,t) = -D\left[\nabla\rho(r,t) + \frac{ze}{k_BT}\rho(r,t)\nabla\Phi(r,t)\right], \quad (2)$$

where $J$ is the current density, $D$ is the constant diffusion coefficient, $\rho$ is the charge density, $z$ is the charge of the ion, $e$ is the elementary charge, $k_B$ is the Boltzmann-constant, $T$ is the temperature and $\Phi$ is the electric potential. To study temporal evolution, this equation has to be combined with the continuity equation, $\nabla J + \frac{\partial \rho}{\partial t} = 0$, which will result the following equation:

$$\frac{\partial \rho}{\partial t} = D\left[\nabla^2\rho + \frac{ze}{k_BT}(\nabla\rho\nabla\Phi + \rho\nabla^2\Phi)\right], \quad (3)$$

The D diffusion coefficient can be derived from the electrical (electron and hole) mobility, using the Einstein-relation, $\mu = \frac{e}{k_BT}D$ [3]. In PTFE, the hole mobility dominates charge transport [28]. However, the accurate data is difficult to find, and, in addition, mobilities on the surface and in the bulk might be different. In our simulation, similarly to other capillary simulations (e.g. [12]), we applied a factor of 100 difference between them, i.e. we used the values $\mu_{surface} = 10^{-12} m^2 V^{-1} s^{-1}$ and $\mu_{bulk} = 10^{-14} m^2 V^{-1} s^{-1}$ (taken from Ref. [28]).

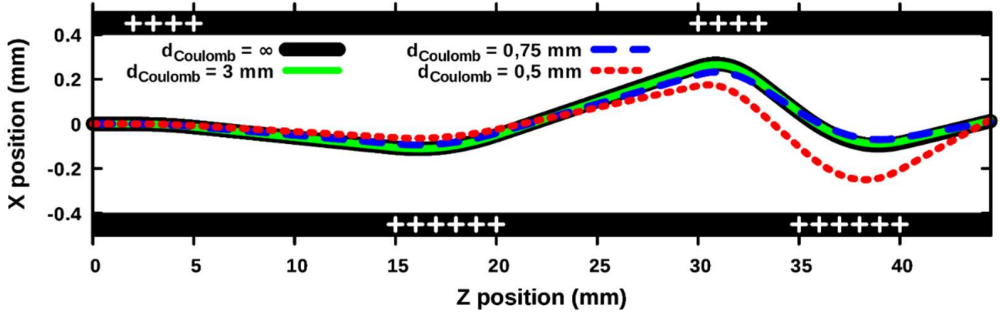

FIG. 7. Effect of the maximization of the range *d* of Coulomb repulsion on the trajectory of a proton.

In our simulation, Eq. (3) was solved by a finite difference method. The capillary wall was approximated by a three dimensional grid, with no grid points inside the bulk material, i.e., the radial dimension contained only the inner and outer surface. The electric potential term is calculated by the sum of n pointlike charges,

$$\Phi(\vec{r},t) = k \cdot \sum_{i=1}^{n} \frac{Q_i}{|\vec{r_i}-\vec{r}|}, \qquad (4)$$

where $Q_i$ represents the charge density in the ith grid point. This is a pseudobulk model, since it does not contain information about the charges inside the bulk, but it does describe the current that flows through the bulk to the direction of the outer surface. As was mentioned, a simulation is always based on a simplified model, but to obtain results as comparable as possible to our earlier experiments, the boundary conditions are given as similar as possible to the experimental work:

(i) along the longitudinal axis we used Dirichlet-type bounds of $\rho(0) = \rho(l) = 0$ and $\Phi(0) = \Phi(l) = 0$, indicating that the sample is mounted into the sample holder (ground potential) at the front and back edges.

(ii) Along the tangential axis, we applied periodic boundary condition (the capillary wall closes into itself).

(iii) Along the radial axis, there were Neumann type bounds of no-flux, i.e. charges can not escape from the surface to the vacuum, and no other charges are captured by the surface, $\frac{\partial}{\partial r} \frac{\partial \rho}{\partial t} = 0$. We note that this approximation certainly contains some neglection, as secondary electrons might affect the charge patches, but we assume that this effect is insignificant. It is based on the assumption that they are re-captured by the generated charge patch itself, or in its vicinity, which is plausible, since the simulated results are in accordance with the experiments.

### F. Optimization

In order to speed up the simulation, several well-known approximations are applied. One of these approximations is the maximization of the range of Coulomb interaction, within which charges interact with each other through Coulomb repulsion. Since the range of Coulomb force is infinite, a cutoff distance beyond which the Coulomb repulsion can be neglected was introduced. This distance, $d_{Coulomb}$, was determined in a series of tests. In the test simulation, the wall of the capillary was charged up "artificially," and single protons were launched, while the cutoff distance of Coulomb repulsion was changed.

Figure 7 shows that by applying a threshold distance of $d_{Coulomb}$ = 3 mm, the trajectory of protons practically equal to the trajectory without applying a cutoff distance, but the run time is significantly shorter. A 3-mm cutoff distance was also successfully applied in Ref. [13].

Another method to accelerate calculations, which is applied in almost every capillary simulation work in the literature, is that every projectile ion of charge *q* (one elementary charge for protons) deposits *Q = Nq* charge on the wall. Applying this crude model, *N* times the lower number of primary projectiles is needed in the simulation. The value of *N* is typically between several thousand [29] and several ten thousand [30] for slow, highly charged, and/or slow ions. For 1-MeV protons test simulations showed that even in the case of $N = 10^6$ the trajectory is only slightly modified, which can be explained by their large electrical rigidity. This is in accordance with the observation published in Ref. [23] by Liu *et al.*, where the authors used $10^6$ for 1 MeV, and claimed that it barely affected the trajectory calculations. Figure 8 shows typical proton trajectories with various charge deposition factors. Here we also call attention to the large difference in the scales between the *x* and *y* axis dimensions. In our simulations, to stay in the safe side, we took the value $N = 10^5$.

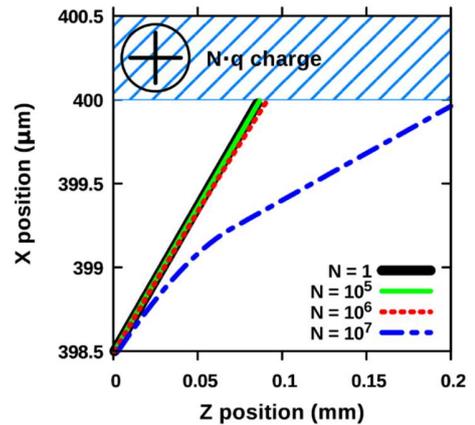

FIG. 8. Typical proton trajectories with various charge deposition factors. The position X = 400 μm is the inner capillary surface.

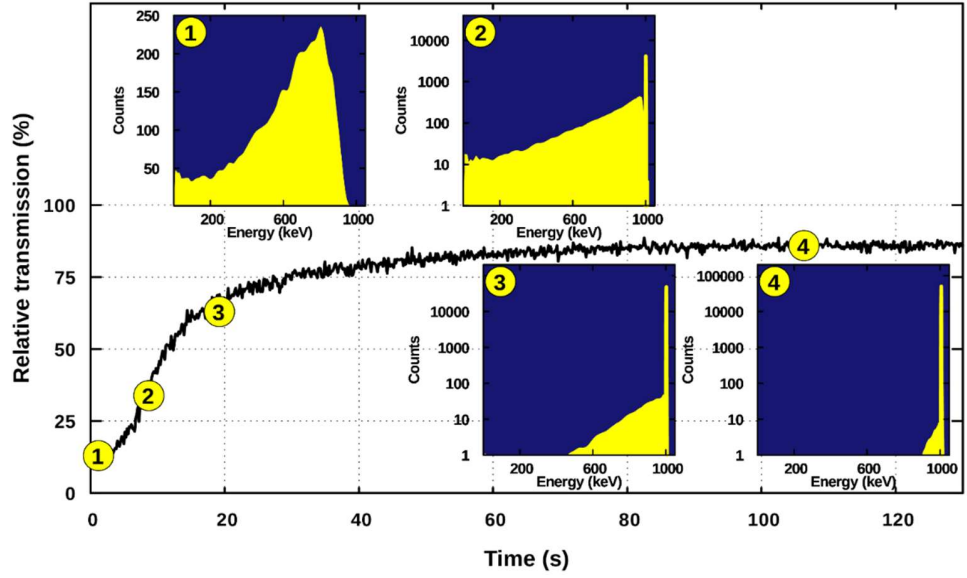

FIG. 9. Calculated relative transmission of the beam as a function of time (or equivalently, inserted charge), and energy spectra of the transmitted particles at different phases of the transmission, namely (1) scattering phase; (2) and (3) mixed phase; (4) guiding phase.

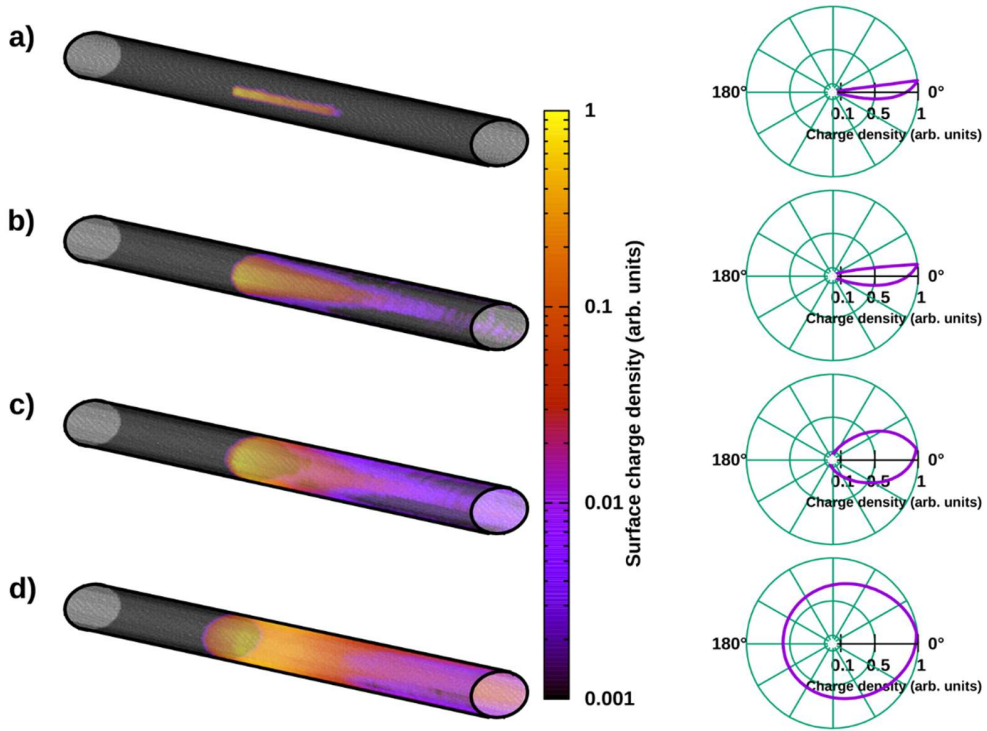

FIG. 10. Charge patch formation and growth in the simulation, in case of 10-pA incident intensity. The beam comes from the left. Left: 3D charge distributions images; right: cross-section images of the charge distribution taken at z 23-mm longitudinal position (the center of the charge patch). (a) Shortly after the simulation starts; (b) at 50% relative transmission; (c) at the saturation of the transmission; (d) during the stabilized transmission, 15 min after saturation.

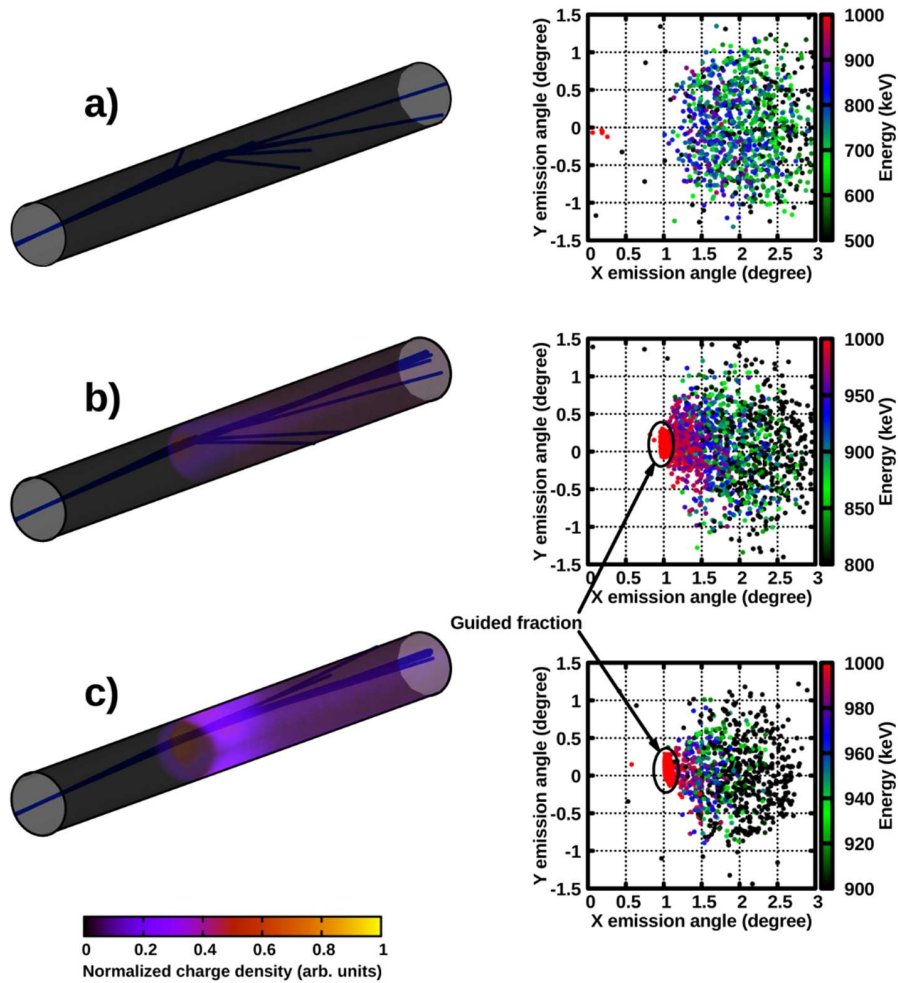

FIG. 11. Simulated particle trajectories (left) and the corresponding angular distributions (right) using 10-pA incident intensity. (a) At the beginning of transmission; (b) during the formation of the charge patch; (c) at stabilized transmission.

## III. RESULTS AND DISCUSSION

### A. Relative transmission and energy spectra

The relative transmission, i.e., the ratio of the transmitted beam relative to the incident beam in percent, Itransmitted Iincident 100%, of the beam and the energy spectrum of the transmitted particles at different phases of the transmission were calculated. Figure 9 shows that the relative transmission starts from around 15%, and immediately starts to gradually increase. Then, at around 90%, it stabilizes. This result agrees well with our experiments [16] as well as our earlier simulations [21]. The evolution of the energy spectra of the transmitted particles also agrees well with our earlier work [19,21], indicating a temporal evolution of the guiding effect:

(1) scattering phase, where the incident beam hits the inner capillary wall, transmission is only possible through multiple small angle forward scattering, and thus the transmitted particles suffer energy loss (see inset 1 of Fig. 9);

(2) mixed phase, where an increasing fraction of the beam is transmitted through the capillary via ion guiding, i.e., without collisions with it and thus without energy loss (see insets 2 and 3 of Fig. 9);

(3) guiding phase, where the majority of the outgoing particles is transmitted through the capillary by ion guiding, without any energy loss (see inset 4 of Fig. 9). Although in the case of ion guiding, the mean value of the kinetic energy as well as its spread is equal to that of the incident beam, inside the capillary they dynamically change according to the formed electric field, as was already observed and published in Ref. [21]: protons approaching the high potential of the charge patches continuously lose their kinetic energy, while protons

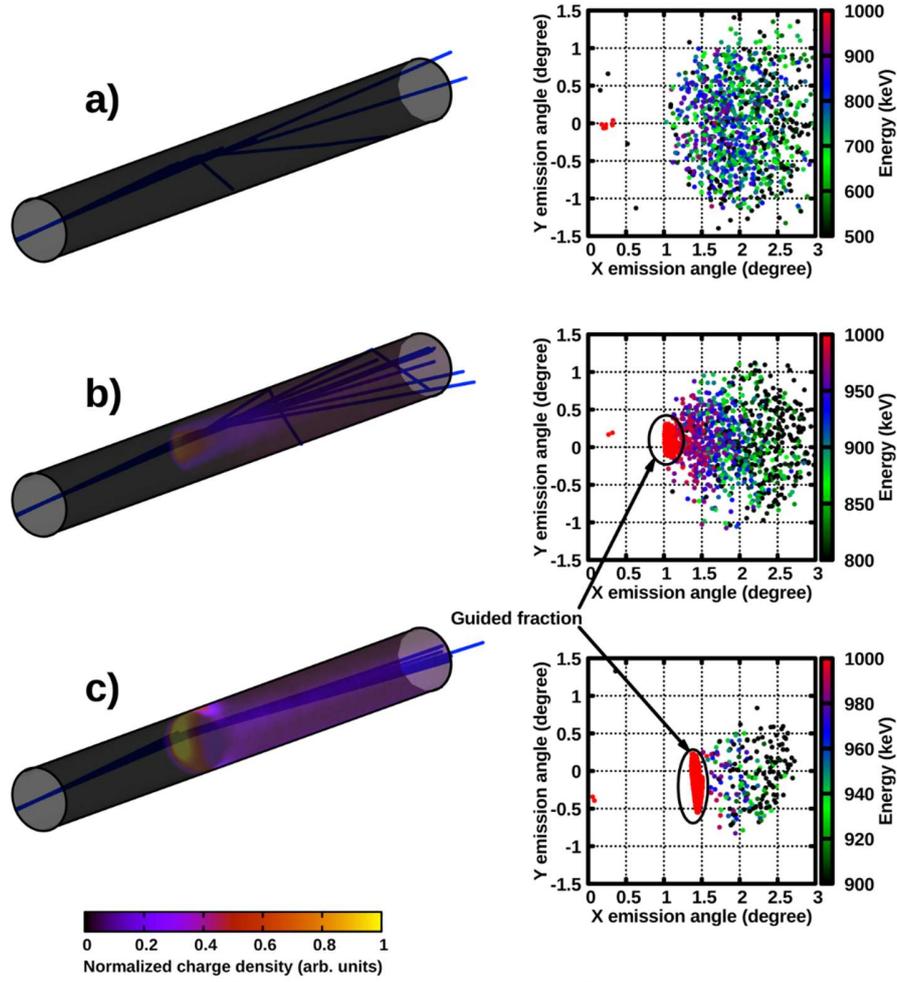

FIG. 12. Simulated particle trajectories (left) and the corresponding angular distributions (right) using 100-pA incident intensity. (a) At the beginning of transmission; (b) during the formation of the charge patch; (c) at stabilized transmission.

leaving the high potential area continuously gain back their initial kinetic energy.

The high level of agreement of these results with our previous works suggests that the implemented methods and formulas in the present work can be considered as a good approximation, and our model is capable of describing the most important processes that are responsible in the formation of the ion guiding phenomenon.

### B. Charge patch formation

The key process in the successful formation of the ion guiding phenomenon is the buildup of the proper charge patches, which generate the guiding electric field. During the simulations, we continuously followed the behavior, i.e., the creation and growth of the charge patches in our system. Figure 10. shows this behavior in the case of 10 pA incident intensity at four different time moments, namely (a) shortly after the simulation is started, and there is no guided transmission yet, (b) at approximately 50% relative transmission, (c) when the transmission is saturating, and (d) during the stable transmission, 15 min after the stabilization. We note that this figure shows only the charge density on the inner capillary wall, but from the calculations we see that the contribution of the outer wall is insignificant, which is due to the large wall thickness of the capillary, i.e., the large bulk resistance. One can see from Fig. 10 that in our present case there is only one major charge patch forming, whose longitudinal extension is large and, in addition, it becomes even larger and larger, thus the produced electric field is broad enough to be able to deflect the beam before hitting the wall. Due to the forward scatterings, charges accumulate in the second half of the capillary practically everywhere, but there is no other strong, well defined patch. It can also be seen that the primary charge patch starts to tangentially flow around the inner wall, resulting in a ring-shaped patch, but spreads slowly

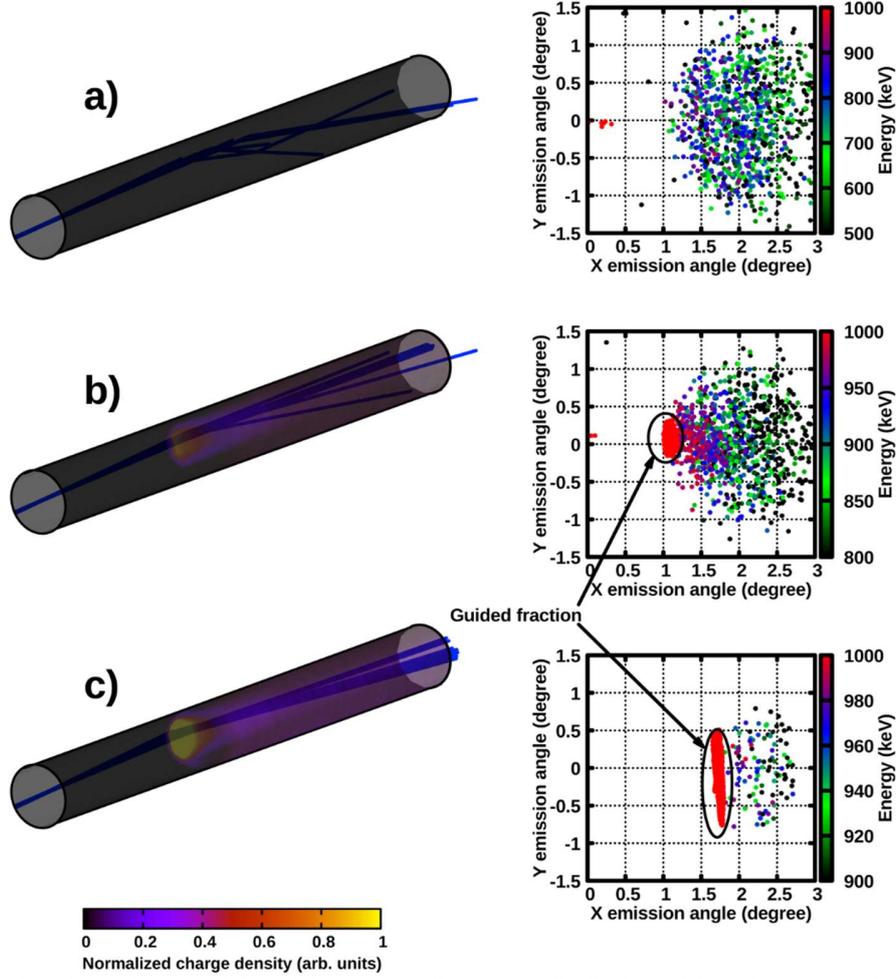

FIG. 13. Simulated particle trajectories (left) and the corresponding angular distributions (right) using 1000-pA incident intensity. (a) At the beginning of transmission; (b) during the formation of the charge patch; (c) at stabilized transmission.

both tangentially and longitudinally due to the large surface resistivity of the material.

### C. Spatial distributions of the transmitted beam

We calculated the spatial distribution of the transmitted beam and its dependence on the level of the relative transmission. The results are shown in Figs. 11–13, for 10-, 100-, and 1000-pA incident beam intensity, respectively. The left-hand columns show trajectories of 20 randomly chosen particles, while the right-hand columns show the corresponding spatial distributions of a large number of particles. It is seen in the (a) panels of the figures that in the first, scattering phase, only particles with energy loss occur in the transmitted beam (the energy spectrum was shown in inset 1 of Fig. 9). These particles are distributed in a wide angular range: more than 95% of the ions are within 1–3° and −1–1° in the horizontal and vertical direction, respectively. We note that in the horizontal direction 1° is the capillary tilt angle, i.e., the direction of the transmitted particles relative to the direction of the capillary axis is within −1–1° in both planes. This is actually the opening angle of the capillary exit seen from the scattering region (i.e., around the middle of the length) of the inner capillary wall. A larger exit angle is possible, though, if a partly deflected proton suffers forward scattering close to the end region of the capillary, or if it is backscattered from the inner wall more than once, but the probability of these scenarios is significantly smaller.

Later, when the relative transmission starts to increase, a small fraction of the beam transmits the capillary without collisions with the inner wall and thus without any significant energy loss. In the (b) panels of Figs. 11–13, they are seen at 1° X angle. In this case, the guided beam fraction is parallel to the axis of the capillary. Forward scattered ions are still present, within the same angular range as they had in the pure scattering phase, but with lower intensity.

Finally, in the guiding phase, the guided fraction dominates the transmission [see the (c) panels of Figs. 11–13]. More-over, this dominance is more pronounced at larger incident intensities, which is explained by the incident beam intensity dependence of the maximum achievable transmission: at larger incident beam intensity, the efficiency of transmission saturates at a higher value [20]. Inversely, in the case of lower incident current, the transmission saturates at a lower relative value due to the discharge and thus the weaker electric field of the formed charge patch. Therefore, a larger fraction of the transmitted beam suffers forward scattering. In the guiding phase, as it is the stabilized state of the guiding process, the final, maximum horizontal deflection can be determined. It is approximately 1°, 1.4°, and 1.7° for 10, 100, and 1000 pA, respectively. This means that in the case of lower incident beam intensity, the outgoing beam leaves the capillary parallel with its axis, while in the case of larger incident intensity, the beam deflection shows a dynamics, and the available maximum deflection angle is larger than the tilting angle of the capillary. This again agrees well with our previous experimental results (Ref. [17] for 10 pA and Ref. [18] for 100 pA).

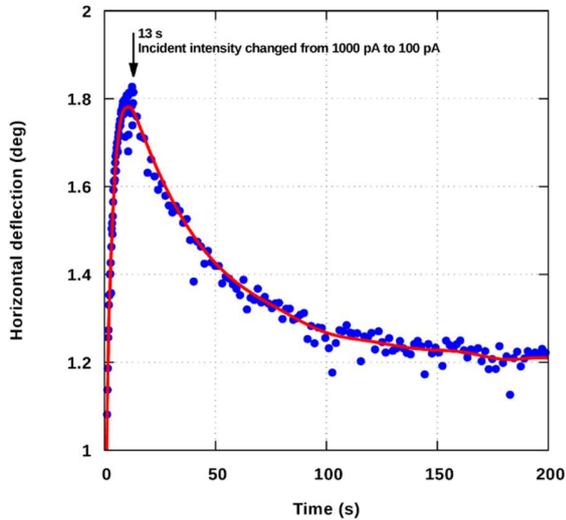

FIG. 14. Horizontal (X) deflection of the beam as a function of time. The incident beam intensity is 1000 pA at the beginning, then it is suddenly changed to 100 pA. The blue dots are the simulated data points, while the red line is shown to guide the eye.

The main objective of the development of simulations is to obtain a tool which is able to examine processes and reliably predict situations which are challenging to observe, due to, for example, the experimental feasibility or the sensitivity of the instruments. In the following we present the case where the simulation starts with 1000-pA incident intensity, and later it suddenly drops to 100 pA. The horizontal deflection of the beam as a function of time is illustrated in Fig. 14. It can be seen that first, when the incident intensity is 1000 pA, the deflection quickly increases, and saturates at around 1.8°. Here, the incident intensity was reduced to 100 pA. From this moment, as expected, the deflection gradually shifts back. Here, the saturation value is around 1.2°, which is lower than in the case when the incident intensity is 100 pA from the beginning of the simulation. This is, however, reasonable, because in this case the total deposited charge is larger, which, after flowing around the inner wall to the opposite side, lowers the deflecting electric field in the capillary. The beam size and the energy spectrum of the transmitted particles also change during the period between the sudden decrease and the establishment of the stable transmission. In the energy spectrum, the number of scattered particles, i.e., the number of particles that suffered energy loss, become higher. At the moment of the sudden drop to 100 pA intensity, approximately 1% of the transmitted particles suffer some energy loss. Later, when the deflection stabilizes, approximately 2.5% suffers energy loss. In addition to this, after the sudden drop, the vertical extension of the beam decreases due to the decrease of the electric field inside the capillary. In the Supplemental Material we provide an animation about the evolution of the beam shape throughout the full simulation presented in Fig. 14 [31].

To illustrate the weight of the guided fraction, Fig. 15 shows the same two-dimensional maps of the transmitted particles as they were shown in the right-hand column of Fig. 12, but instead of the particle energy, the color scale corresponds to the current density (i.e., the number of particles per unit area). It becomes clear that in the guiding phase the majority of the transmitted particles belong to the guided beam fraction: the intensity density of the scattered fraction is two to three orders of magnitude smaller than that of the guided fraction. It is also evident that some scattered particles are present, although in our former experiments (in Refs. [17,18]) we were not able to detect them. According to our simulation the explanation is as follows: The intensity of the scattered fraction was lower than the detection limit of our experimental setup.

### D. Effect of the microbeam size

As shown in Fig. 2, the incident beam divergence is independent of the beam intensity, but the beam size does depend on it: larger beam intensity results in a larger beam spot size (FWHM). This is reasonable and acceptable, because the intensity is adjusted only by the object apertures. Some test simulations showed that this change in the beam size has no observable effect on the scattering and guiding process. Although the spot size changes almost an order of magnitude by increasing the beam intensity from 10 to 1000 pA, it is still pointlike with respect to the capillary dimensions therefore the size of the formed charge patch on the wall will have the same dimensions. Our estimation showed that increasing the spot size from 1 to 10 μm results in a less than 5% increase of the primary charge patch. On the other hand, the beam divergence plays an important role, because it strongly affects the longitudinal extension of the charge patch. Moreover, by increasing the divergence, the maximum energy originating from the transversal velocity component of the beam, which must be compensated by the patch, is also increasing. Inversely, in the case of a larger divergence, the same potential of the capillary will not be able to deflect all incident protons. However, the beam divergence does not necessarily change with the intensity, and therefore, it is unchanged in the simulation. In conclusion, the change in the beam

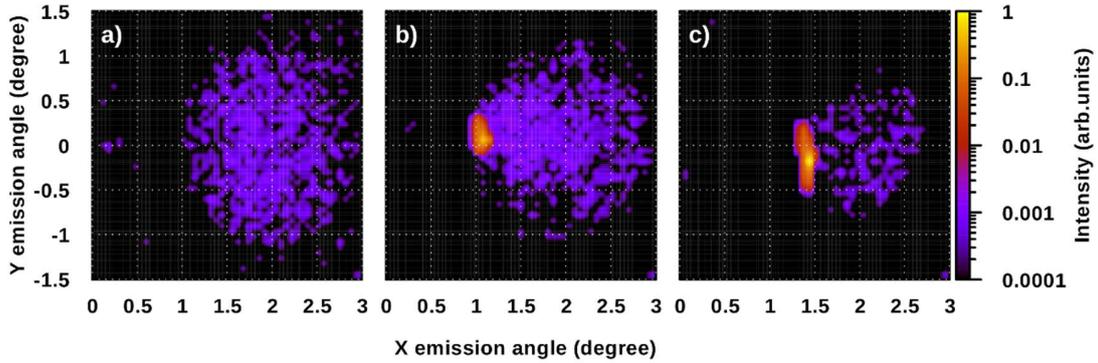

FIG. 15. Simulated angular distribution of the transmitted beam, using 100 pA incident intensity. (a) At the beginning of transmission; (b) during the formation of the charge patch; (c) at stabilized transmission.

size did not influence noticeably the results of the present work.

### E. Beam focusing and defocusing

Finally, we present the change of the shape of the transmitted beam. Figure 16 shows the horizontal (X) and vertical (Y) angular distributions of the transmitted beam compared to that of the incident beam. These are full line profiles, for which all rows or columns of Fig. 15(c) were integrated in order to obtain horizontal and vertical profiles, respectively. In the X direction, the beam is focused: the full opening angle of the transmitted beam is approximately 0.15°, instead of the initial 0.46°. However, in the Y direction, the beam is defocused: although it does not have sharp edges, it can be seen that the full opening angle of the transmitted beam is much wider than that of the incident beam, which is approximately 0.16°. This phenomenon can be attributed to a quadrupolelike focusing effect, which is caused by the quadrupole moment of the deposited charge patch, and which was experimentally observed in Ref. [18].

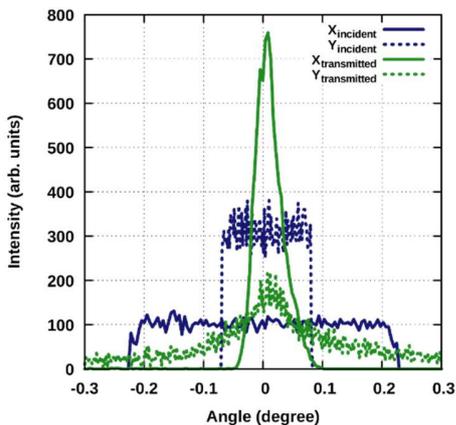

FIG. 16. Angular distribution of the incident and transmitted beam, using 100-pA incident intensity. The curves are zero centered and normalized to each other, to be able to easily compare them.

### IV. CONCLUSIONS

We have presented computer simulations for the spatial and temporal evolution of a 1-MeV proton microbeam transmitted through a single, insulating macrocapillary. The simulations were performed with a cylindrical-shape poly (tetrafluoroethylene) capillary with the length of 45 mm and the inner diameter of 800 μm. The axis of the capillary was tilted to 1° relative to the axis of the incident beam, which ensured geometrical nontransparency. The results were obtained by a three-dimensional computer simulation code that employs both stochastic (Monte Carlo) and deterministic methods. The simulation successfully combines different random sampling methods according to distributions provided by SRIM and WINTRAX, with the numerical solution of the governing equations for following the trajectory of the projectile protons and for describing the field-driven charge redistribution on the surface and in the bulk of the insulating material. The results of the simulation describe well all of our previous experimental observations, indicating the functionality and reliability of the applied methods. We found that at different phases of the beam transmission, different atomic processes result in the change of the beam shape. At the beginning, small angle multiple forward scatterings dominate the transmission, resulting in an outgoing beam into a wide angular range and in a wide energy window. Later, a guided fraction of the beam appears in a small, well-separated beam spot, with kinetic energy equal to that of the incident beam. Finally, a stabilized, guided transmission forms, where the dominant fraction of the beam transmits the capillary with ion guiding, it is concentrated into a small beam spot, and a quadrupolelike focusing effect is identified.

### ACKNOWLEDGMENTS

This work was supported by the RADIATE project under Grant Agreement No. 824096 from the EU Research and Innovation Programme HORIZON 2020. This work was also supported by Project International de Coopération Scientifique Hongrie 2018 Grant No. 245 358 and by the Balaton 2018 program, Projects No. 40301VK, No. NKM-116/2018, and No. NKM-38/2019.